\documentclass[conference]{IEEEtran}

\usepackage{hyperref}
\usepackage{amsmath}
\usepackage{amssymb}
\usepackage{graphicx}
\usepackage{epstopdf}
\usepackage{upgreek}

\begin{document}
\title{Transverse forces in planar symmetric dielectric laser-driven accelerators}

\author{\IEEEauthorblockN{R. Joel England}
\IEEEauthorblockA{SLAC National Accelerator Laboratory\\ Menlo Park, CA 94025, USA \\
Email: england@slac.stanford.edu}
\and
\IEEEauthorblockN{Alexander Ody}
\IEEEauthorblockA{Department of Applied Physics\\ Stanford University \\ Stanford, CA 94305, USA \\}
\and
\IEEEauthorblockN{Zhirong Huang}
\IEEEauthorblockA{SLAC National Accelerator Laboratory \\ Menlo Park, CA 94025, USA}}


%Telephone: (800) 555--1212, Fax: (888) 555--1212}


\maketitle

\begin{abstract}
The use of dielectric microstructures driven by solid state lasers to accelerate charged particles or to transversely deflect them is a growing area of scientific interest with an international collaboration of researchers working to develop this concept. Many experimental efforts and new designs use a planar symmetric geometry with sub-micron apertures for the particle beam. We provide a general theoretical framework for the electromagnetic fields in this type of geometry, including derivation of the TE and TM modes supported, and examine the transverse force components exerted on the beam, which may be used for focusing or for deflection of the particles. 
\end{abstract}


% \fi
\IEEEpeerreviewmaketitle

\section{Introduction}
\label{sec:intro}
    
The use of micron-scale photonic structures powered by solid state lasers to accelerate charged particles is a growing area of scientific interest with a substantial international research effort now underway \cite{wootton_towards_2017}. Although dielectric laser-driven accelerators (DLAs) have been proposed using a variety of 1D, 2D and 3D geometries \cite{plettner_proposed_2006,lin:2001,cowan:2008,wu:pac2013} most structure designs currently under experimental testing are based on a planar-symmetric geometry that is periodic along the particle beam propagation direction (here taken to be the $z$ axis) and invariant in one orthogonal coordinate ($x$) with a narrow sub-wavelength vacuum channel in the other transverse dimension ($y$). The fields for various specific cases of this sort have been previously considered (e.g. \cite{plettner_proposed_2006,plettner:2009,chen:2018,niedermayer:focusing:2018}). Here we consolidate these results to provide a concise general derivation of all non-vanishing field components for both transverse electric (TE) and transverse magnetic (TM) modes for open and enclosed structures under both single-sided and dual-driven laser illumination for either in-phase or $\pi$ out-of-phase lasers, and the corresponding force equations under rotation of the electron axis by an arbitary angle. These relations can be used to identify useful configurations for laser-driven acceleration or for deflection in either transverse dimension. 
   
%%%%%%%%%%%%%%%%%%%%%%%%%%%%%%%%%%%%%%%%%%%%%%%%%%%%%%%%%%%%%%%%%%%%%%%%%%%%%%%%%%
%%%%%%%%%%%%%%%%%%%%%%%%%%%%%%%%%%%%%%%%%%%%%%%%%%%%%%%%%%%%%%%%%%%%%%%%%%%%%%%%%%
%%%%%%%%%%%%%%%%%%%%%%%%%%%%%%%%%%%%%%%%%%%%%%%%%%%%%%%%%%%%%%%%%%%%%%%%%%%%%%%%%%
\section{Formulation of Fields}
\label{sec:theory}

\subsection{General Field Relations}

\begin{figure}
\includegraphics[width=\columnwidth]{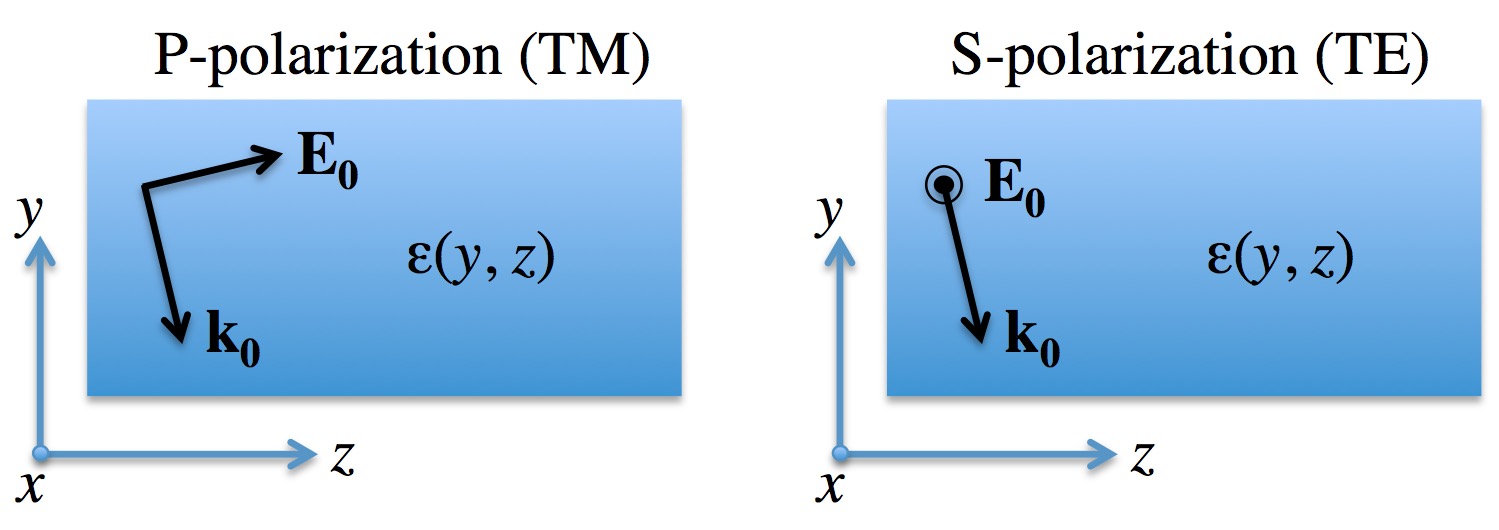}%
\caption{Schematic of coordinate system and geometry for P and S polarization for a linear dielectric that is uniform in the $x$ (out of page) coordinate.}%
\label{fig:geometry}
\end{figure}

In the absence of free sources and assuming harmonic time dependence $e^{-i \omega t}$, Maxwell's equations have the form
\begin{equation}
\nabla \cdot \textbf{D} = \nabla \cdot \textbf{B} = 0  \ , \ \nabla \times \textbf{E} = i \kappa \textbf{B} \ , \ \nabla \times \textbf{H} = - i \kappa \textbf{D}
\label{eqn_maxwell}
\end{equation}
where $\kappa = \omega / c$ is the free space wavenumber and the field quantities are assumed to have only a spatial dependence. We consider a dielectric medium in which the magnetic permeability $\mu = 1$ everywhere but with a dielectric constant $\hat{\epsilon}(\textbf{r})$ having spatial dependence.  In this case the electric displacement field and magnetic induction are given by $\textbf{D} = \hat{\epsilon}\  \textbf{E}$ and $\textbf{B} = \textbf{H}$. Hence, we can derive from Eqs. (\ref{eqn_maxwell}) the following wave equation for the electric field
\begin{equation}
\nabla^2 \textbf{E} - \nabla (\nabla \cdot \textbf{E}) = - \kappa^2 \textbf{D}
\label{eqn:waveeqn}
\end{equation}
where we have used the identity $\nabla \times \nabla \times \textbf{A} = \nabla (\nabla \cdot \textbf{A}) - \nabla^2 \textbf{A}$. Constraining the problem to a system that is translationally invariant in the $x$ coordinate, two orthogonal polarizations may then be defined relative to the plane of $y$ and $z$ wherein there is variation of the fields, as shown in Fig. \ref{fig:geometry} and Table \ref{table}. We may call these S and P or conversely transverse electric (TE) and transverse magnetic (TM) modes, where the term ``transverse" is used in reference to the electron axis ($z$). Once a solution for $\textbf{E}$ is in hand, the magnetic field $\textbf{B}$ may then be immediately obtained from the third of Eqs. (\ref{eqn_maxwell}). 

\begin{table}
\centering
\caption{
\label{table}
Polarizations and Nonvanishing Field Components
}
\begin{tabular}{|c|c|c|c|}
  \hline
  Polarization  & Mode & Nonzero E & Nonzero B \\
  \hline
S & TE &  $E_x$   & $B_{y,z}$ \\
P  & TM & $E_{y,z}$  & $B_x$ \\
  \hline
\end{tabular}
\end{table}

By the Floquet Theorem, the solutions to Maxwell's equations subject to periodic boundary conditions along coordinate $z$ with periodicity $\textbf{u} = \lambda_p \hat{\textbf{z}}$ satisfy $\textbf{E} (\textbf{r} + \textbf{u}) = \textbf{E} (\textbf{r}) e^{i \psi}$ and $\textbf{B} (\textbf{r} + \textbf{u}) = \textbf{B} (\textbf{r}) e^{i \psi}$, where $\psi$ is the phase shift from one cell to the next. For traveling wave solutions with the fundamental Bloch wavenumber $\textbf{k}_0$, this implies $e^{i \textbf{k}_0 \cdot \textbf{u}} = e^{i \psi}$ or $\psi = k_0 \lambda_p$. The electric field can thus be decomposed into a Fourier series with longitudinal wavenumbers $k_n = k_0 + n k_p$ where $k_0 = \psi / \lambda_p$ and $k_p = 2 \pi / \lambda_p$. If these fields are excited by an incident plane wave $\textbf{E}_0 =  \hat{\textbf{e}} E_0 e^{i \textbf{k}_0 \cdot \textbf{r} - i \omega t}$, where $\textbf{k}_0 = \kappa \sqrt{\epsilon_i} \hat{\textbf{n}}$, then the phase advance $\psi$ per period is given by the projection of the incident plane wave onto the fundamental periodicity. Here $\epsilon_i$ is the dielectric constant in the region from which the plane wave is incident. Then, $k_0 = | \textbf{k}_0 \cdot \textbf{u} | = \kappa \sqrt{\epsilon_i} \cos \theta$, where $\theta$ is the incidence angle of the laser ($\cos \theta = \hat{\textbf{n}} \cdot \hat{\textbf{z}}$).  The phase velocity of the $n$'th space harmonic, normalized to the speed of light is thus $\beta_n = \kappa / k_n$. If we set this to match the particle velocity $\beta$ we thus obtain the phase matching relation
\begin{equation}
\frac{\beta n \lambda}{\lambda_p} = 1- \beta \sqrt{\epsilon_i} \cos \theta \  ,
\label{eqn:phasematch}
\end{equation}
which for normal incidence ($\theta = \pi / 2$) assumes the familiar form $\lambda_p = \beta n \lambda$.

\subsection{Fields for Single Drive Laser}

\begin{figure}
\includegraphics[width=\columnwidth]{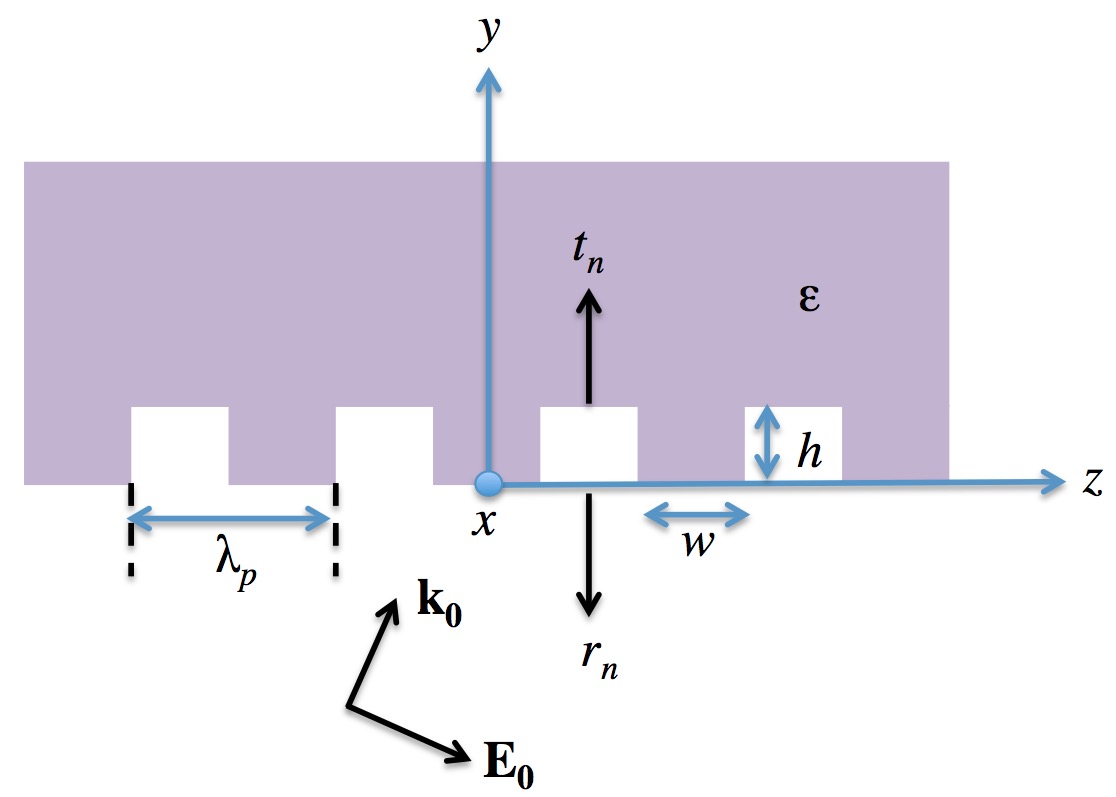}%
\caption{Case of a periodic open structure occupying the upper half-plane ($y>0$) with laser incident from below.}%
\label{fig:halfplane}
\end{figure}

We consider first the case where the periodic structure occupies the half-space $y > 0$ and is excited by an incident plane wave propagating from the vacuum region $y<0$. This scenario is depicted in Fig. \ref{fig:halfplane}. It is shown in \cite{pilozzi:1996} that the solutions to Eq. (\ref{eqn:waveeqn}) in the vacuum region ($y<0$) have the following form for S-polarization
\begin{equation}
E_x (y,z) = E_0 \sum_n [ \delta_{n,0} e^{i \Lambda_n y} + r_n e^{-i \Lambda_n y}] e^{i k_n z},
\label{eqn:spol}
\end{equation}
and for P-polarization
\begin{equation}
\begin{aligned}
&E_y (y,z) = - E_0 \sum_n \frac{k_n}{\Lambda_n}[ \delta_{n,0} e^{i \Lambda_n y} - r_n e^{-i \Lambda_n y}] e^{i k_n z}, \\
&E_z (y,z) = E_0 \sum_n [ \delta_{n,0} e^{i \Lambda_n y} + r_n e^{-i \Lambda_n y}] e^{i k_n z}.
\end{aligned}
\label{eqn:ppol}
\end{equation}
Here $\Lambda_n = \sqrt{\kappa^2 - k_n^2}$, $\delta_{n,m}$ is the Kronecker delta, and $r_n$ is the reflection coefficient of the incident wave for the $n$'th harmonic. This form is general, and direct solution for the $r_n$ requires specification of the geometry. The geometry of Fig. \ref{fig:halfplane} (i.e. a rectangular grating) is chosen as an example but other geometrical configurations (see Fig. \ref{fig:dualstructure}(b)) are also possible. We further note that the Kronecker delta term ($n=0$) in Eq. (\ref{eqn:ppol}) corresponds to the incident plane wave $\textbf{E}_0 $ with the association $\textbf{k}_0 = \Lambda_0 \hat{\textbf{y}} + k_0 \hat{\textbf{z}}$.

Now consider the scenario of two such periodic structures mirrored about the $x$-$z$ plane and separated by a vacuum gap of width $g$, as illustrated in Fig. \ref{fig:dualstructure}(a). Under excitation by a single plane wave $\textbf{E}_0$ from below, by extension of the results of Eqs. (\ref{eqn:spol}-\ref{eqn:ppol}), the fields in the vacuum region will have the following form for S-polarization (TE)

\begin{equation}
\begin{aligned}
&E_x = E_0 \sum_n [ a_n e^{\Gamma_n y} + b_n e^{-\Gamma_n y}] e^{i k_n z}, \\
&B_y  = E_0 \sum_n \frac{k_n}{\kappa}[ a_n e^{\Gamma_n y} + b_n e^{-\Gamma_n y}] e^{i k_n z}, \\
&B_z = i E_0 \sum_n \frac{\Gamma_n}{\kappa} [ a_n e^{\Gamma_n y} - b_n e^{-\Gamma_n y}] e^{i k_n z}.
\end{aligned}
\label{eqn:singleTE}
\end{equation}

and for P-polarization (TM)

\begin{equation}
\begin{aligned}
&E_y = - i E_0 \sum_n \frac{k_n}{\Gamma_n}[ a_n e^{\Gamma_n y} - b_n e^{-\Gamma_n y}] e^{i k_n z}, \\
&E_z = E_0 \sum_n [ a_n e^{\Gamma_n y} + b_n e^{-\Gamma_n y}] e^{i k_n z}, \\
&B_x = i E_0 \sum_n \frac{\kappa}{\Gamma_n} [ a_n e^{\Gamma_n y} - b_n e^{- \Gamma_n y}] e^{i k_n z}. 
\end{aligned}
\label{eqn:singleTM}
\end{equation}
where the complex coefficients $a_n$ and $b_n$ account for the accumulated reflections with associated phase shifts for the various space harmonics within the channel. We also clarify that these coefficients may have different amplitudes and phases for the two modes (TE vs. TM). Since we are interested in modes confined in $y$ and propagating in $z$ we define for convenience a real-valued transverse decay constant $\Gamma_n \equiv i \Lambda_n$. The magnetic field components are obtained by way of 
\begin{equation}
i \kappa \textbf{B}= \nabla \times \textbf{E}=\left\{
\begin{array}{ccc}
 \left[\frac{\partial E_x}{\partial z}\hat{\textbf{y}}-\frac{\partial E_x}{\partial y}\hat{\textbf{z}}\right] & ; & S (\text{TE}) \\
 \left[\frac{\partial E_z}{\partial y}-\frac{\partial E_y}{\partial z}\right]\hat{\textbf{x}} & ; & P (\text{TM}) \\
\end{array}
 \right.
\label{eqn:delcrossE}
\end{equation}
and for completeness, we reiterate the following definitions: $k_n = k_0 + n k_p$, $\Gamma_n = \sqrt{k_n^2 - \kappa^2}$, $k_p = 2 \pi / \lambda_p$, $k_0 = \kappa \sqrt{\epsilon} \cos \theta$, and $\kappa = \omega / c$.

\subsection{Fields for Dual Drive Lasers}

\begin{figure}
\includegraphics[width=\columnwidth]{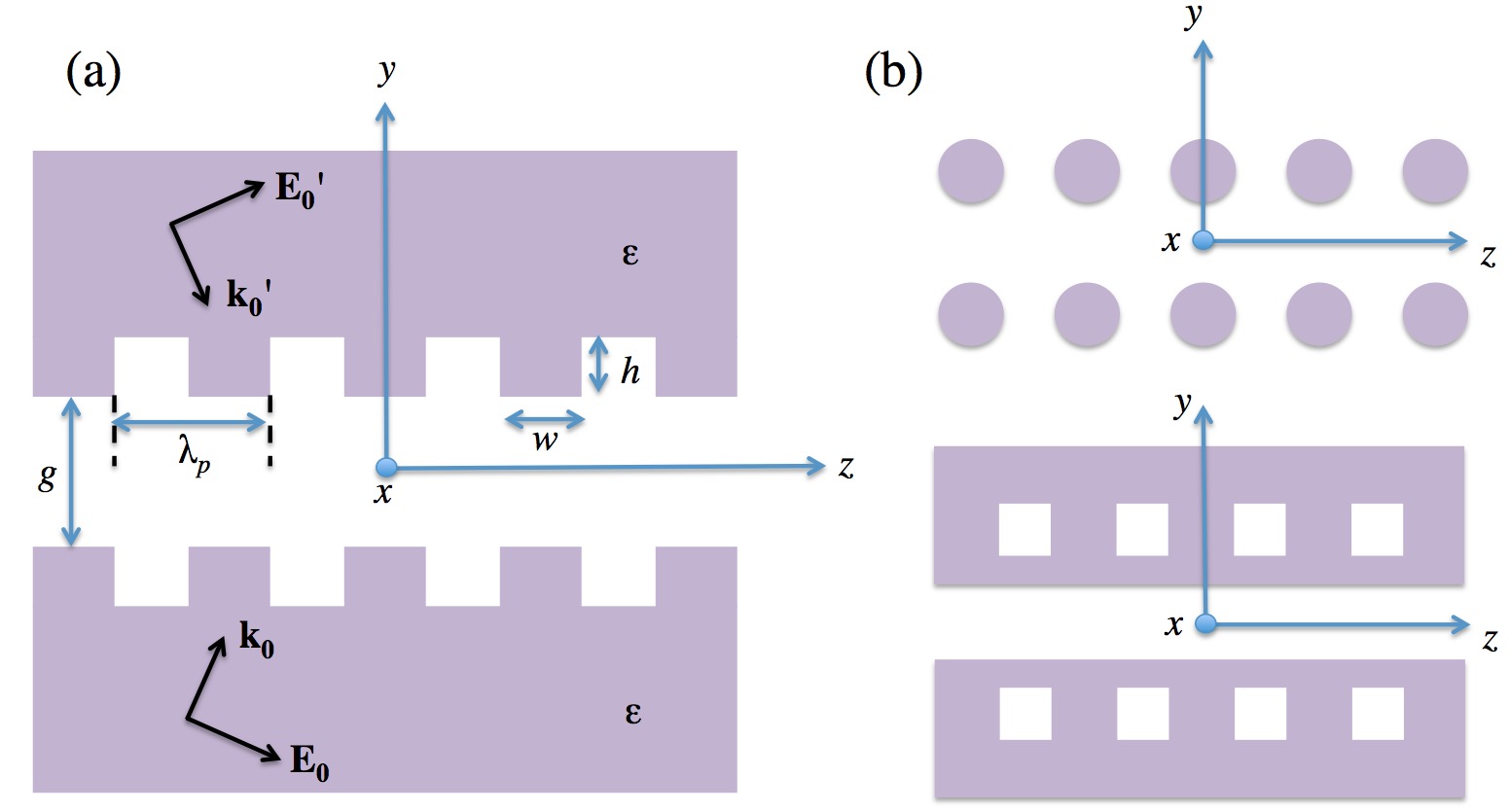}%
\caption{(a) Case of a dual-sided periodic structure (periodicity $\lambda_p$) invariant in $x$ and with mirror symmetry about the $x$-$z$ plane. (b) Alternative geometries that have been considered in recent structure designs.}%
\label{fig:dualstructure}
\end{figure}

If we now allow for a counter-propagating excitation $\textbf{E}_0'$ illuminating the structure from the opposite site (i.e. propagating from the region $y>0$), then we have an additional set of field components ($\textbf{E}'$, $\textbf{B}'$) induced by the counter-propagating wave. These are of the same form as Eqs. (\ref{eqn:singleTE},\ref{eqn:singleTM}) but with corresponding coefficients $a_n'$ and $b_n'$ and potentially different amplitudes and wavenumbers ($E_0'$, $\Lambda_n'$, and $k_n'$). However, if $\textbf{E}_0'$ has the same magnitude, incidence angle, and polarization as $\textbf{E}_0$, then by the mirror symmetry of Fig. \ref{fig:dualstructure}, $b_n' = a_n$ and $a_n' = b_n$. In combining the two solutions we can choose to add or subtract them, corresponding to lasers that are either perfectly in-phase or $\pi$ out-of-phase with each other. We thus obtain the following forms for the case of S-polarization (TE)
\begin{equation}
\begin{aligned}
&E_x=E_0\sum _n c_n^\pm \left\{
\begin{array}{c}
 \cosh \left(\Gamma _ny\right) \\
 \sinh \left(\Gamma _ny\right) \\
\end{array}
\right\}e^{i k_nz}\text{  }  \\
&B_y=  E_0\sum _n  \frac{k_n}{\kappa} c_n^\pm \left\{
\begin{array}{c}
 \cosh \left(\Gamma _ny\right) \\
 \sinh \left(\Gamma _ny\right) \\
\end{array}
\right\}e^{i k_nz}\text{  } \\
&B_z=i  E_0\sum _n   \frac{\Gamma _n}{\kappa} c_n^\pm \left\{
\begin{array}{c}
 \sinh \left(\Gamma _ny\right) \\
 \cosh \left(\Gamma _ny\right) \\
\end{array}
\right\}e^{i k_nz}\text{  }
\end{aligned}
\label{eqn:dualTE}
\end{equation}
and P-polarization (TM),
\begin{equation}
\begin{aligned}
&E_y=-i  E_0\sum _n  \frac{k_n}{\Gamma _n} c_n^\pm \left\{
\begin{array}{c}
 \sinh \left(\Gamma _ny\right) \\
 \cosh \left(\Gamma _ny\right) \\
\end{array}
\right\}e^{i k_nz}\text{  }\\
&E_z= E_0\sum _n  c_n^\pm \left\{
\begin{array}{c}
 \cosh \left(\Gamma _ny\right) \\
 \sinh \left(\Gamma _ny\right) \\
\end{array}
\right\}e^{i k_nz}\text{  }\\
&B_x=i E_0\sum _n  \frac{\kappa}{\Gamma _n} c_n^\pm \left\{
\begin{array}{c}
 \sinh \left(\Gamma _ny\right) \\
 \cosh \left(\Gamma _ny\right) \\
\end{array}
\right\}e^{i k_nz}
\end{aligned}
\label{eqn:dualTM}
\end{equation}
where we define $c_n^\pm \equiv 2 (a_n \pm b_n)$ and the upper (lower) lines correspond to in-phase ($\pi$ out-of-phase) incident fields. Here, we see that the exponential terms in Eqs. (\ref{eqn:singleTE},\ref{eqn:singleTM}) give rise to hypbolic functions ($\cosh$, $\sinh$) depending upon whether they are added or subtracted. We note that the desired mode for acceleration is the in-phase TM mode (upper line of Eq. \ref{eqn:dualTM}). 

The hyperbolic cosine dependence can be seen to approach a transversely uniform field in the limits where either the vacuum gap is small compared to the transverse exponential decay term ($g \ll \Gamma_n^{-1}$) and/or the phase velocity approaches the speed of light ($\beta_n = 1$). We can regard this limiting case by taking the Taylor series of Eqs. (\ref{eqn:dualTE},\ref{eqn:dualTM}) to zero'th order. For a single harmonic, which we call $n = r$ assumed to be synchronous with the desired particle velocity, we thereby obtain 
\begin{equation}
\begin{aligned}
&E_z = - i \hat{E}_0  e^{i k_r z} \  \\
&E_y = - \hat{E}_0  k_r y e^{i k_r z} \ \\
&B_x = \hat{E}_0 k_r \beta_r y e^{i k_r z}\ 
\end{aligned}
\label{eqn:linearized}
\end{equation}
where we have absorbed the constant $2 (a_r + b_r)$ into $\hat{E}_0$ and multipled by an arbitrary phase constant $-i$. This linearized form is useful for simple particle dynamics calculations in generic planar-symmetric structures.

\subsection{Calculation of Transverse Forces}

We can simplify the field equations by defining some functions that separately represent the spatial and geometrical dependences. Let the function $\text{hyp}^\pm(x) \equiv e^x \pm e^{-x}$. Then define
\begin{equation}
\textbf{f}^{\pm }(x)\equiv \left(
\begin{array}{c}
 \text{hyp}^{\pm }(x) \\
 \text{hyp}^{\mp }(x) \\
 \text{hyp}^{\pm }(x) \\
\end{array}
\right)\text{   };\text{   }\textbf{G}_n^{\pm }(\textbf{r}) \equiv  \textbf{f}^{\pm }\left(\Gamma _ny\right)e^{i \textbf{k}_n\cdot \textbf{r}}\text{   }
\end{equation}
Then Eqs. (\ref{eqn:dualTE},\ref{eqn:dualTM}) can be reduced to the simplified form:
\begin{equation}
\begin{aligned}
&\textbf{E}=E_0\sum _n c_n^{\pm }\textbf{G}_n^{\pm }(\textbf{r})\circ \pmb{\mathcal{E}}_n \\
&\textbf{B}=E_0\sum _n c_n^{\pm }\textbf{G}_n^{\mp}(\textbf{r})\circ \pmb{\mathcal{B}}_n
\end{aligned}
\end{equation}
where $\circ$ is the Hadamard product, and $\pmb{\mathcal{E}}_n$ and $\pmb{\mathcal{B}}_n$ have the following forms for the two mode types:
\begin{equation}
\begin{aligned}
&\pmb{\mathcal{E}}_n^{\text{TE}}=\left(
\begin{array}{c}
 1 \\
 0 \\
 0 \\
\end{array}
\right)  
&\pmb{\mathcal{B}}_n^{\text{TE}}=\left(
\begin{array}{c}
 0 \\
 \left.k_n\right/\kappa  \\
 i \left.\Gamma _n\right/\kappa  \\
\end{array}
\right) \text{   } \\
&\pmb{\mathcal{E}}_n^{\text{TM}}=\left(
\begin{array}{c}
 0 \\
 -i k_n/\Gamma _n \\
 1 \\
\end{array}
\right) 
&\pmb{\mathcal{B}}_n^{\text{TM}}=\left(
\begin{array}{c}
 i \kappa \left/\Gamma _n\right. \\
 0 \\
 0 \\
\end{array}
\right) \text{   }
\end{aligned}
\end{equation}

The resulting force on a particle of charge $q$ traveling on the $z$-axis with velocity $v = \beta c$ is then given by
\begin{equation}
\textbf{F}=q E_0\sum _n c_n^{\pm }\textbf{G}_n^{\pm }(\textbf{r})\circ \textbf{Q}_n
\label{eqn:force}
\end{equation}
where $\textbf{Q}_n \equiv \pmb{\mathcal{E}_n}+\beta\hat{\textbf{z}} \times \pmb{\mathcal{B}}_n$. This gives rise to the following forms for the $\textbf{Q}_n$ vector for the TE and TM cases:
\begin{equation}
\textbf{Q}_n^{\text{TE}}=\left(
\begin{array}{c}
 1-\frac{\beta }{\beta _n} \\
 0  \\
 0 \\
\end{array}
\right)\text{  },\text{  }
\textbf{Q}_n^{\text{TM}}=\left(
\begin{array}{c}
 0 \\
 i \frac{k_n}{\Gamma _n}\left(\beta  \beta _n-1\right)  \\
 1 \\
\end{array}
\right)
\label{eqn:Qunrotated}
\end{equation}
These forms demonstrate that in the limit where the particle is both relativistic and matched to the phase velocity of the wave ($\beta = \beta_n \rightarrow 1$), the TM mode provides an accelerating force in $z$ while the transverse force components vanish. Similarly, for the TE mode, in this limit the corresponding $x$ force vanishes. Hence there is no synchronous solution for deflection of relativistic beams. 

\section{Synchronous Force in the Rotated Frame}

\begin{figure}
\includegraphics[width=\columnwidth]{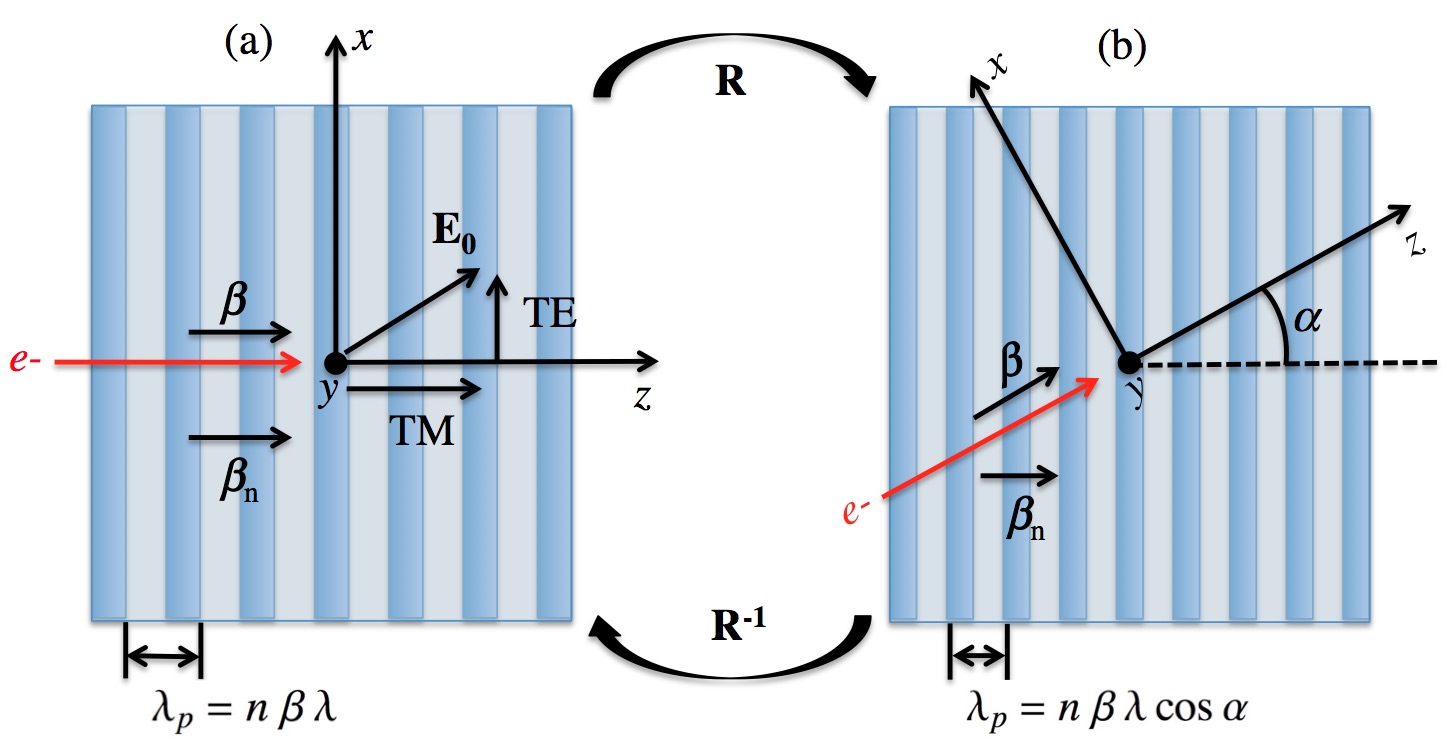}%
\caption{Illustration showing (a) original unrotated geometry with beam axis and (b) after rotation of the coordinates by angle $\alpha$ via the rotation matrix $\textbf{R}$.}%
\label{fig:rotated}
\end{figure}

The solution proposed by Plettner and Byer \cite{plettner:2008} to achieve synchronous deflection is to rotate the beam axis about $y$ by an angle $\alpha$, as illustrated in Fig. \ref{fig:rotated}. This can be represented by a rotation matrix  
\begin{equation}
\textbf{R}=\left(
\begin{array}{ccc}
 \cos  \alpha  & 0 & -\sin  \alpha  \\
 0 & 1 & 0 \\
 \sin  \alpha  & 0 & \cos  \alpha  \\
\end{array}
\right).
\end{equation}
The form for the force $\textbf{F}$ represented in the rotated beam coordinates (keeping $z$ along the particle direction) is still given by Eq. (\ref{eqn:force}), except that we make the replacements
\begin{equation}
\textbf{Q}_n \rightarrow \textbf{R} \cdot \pmb{\mathcal{E}_n}+\beta\hat{\textbf{z}} \times \textbf{R} \cdot \pmb{\mathcal{B}}_n \ \  ,\ \  \textbf{k}_n \rightarrow \textbf{R} \cdot \textbf{k}_n \ .
\label{eqn:transform}
\end{equation}
Under the transformation of Eq. (\ref{eqn:transform}), the $\textbf{Q}_n$ vectors for the TE and TM modes are found to be
\begin{equation}
\begin{aligned}
&\textbf{Q}_n^{\text{TE}}=\left(
\begin{array}{c}
 \cos \alpha-\frac{\beta }{\beta _n} \\
-i \frac{\Gamma_n}{\kappa} \beta \sin \alpha  \\
 \sin \alpha \\
\end{array}
\right), \\
&\textbf{Q}_n^{\text{TM}}=\left(
\begin{array}{c}
 - \sin \alpha \\
 i \frac{k_n}{\Gamma _n}\left(\beta  \beta _n \cos \alpha -1\right)  \\
 \cos \alpha \\
\end{array}
\right) .
\end{aligned}
\end{equation}

The resonant velocity $\pmb{\beta}_n$ is still in the direction of $\textbf{k}_n$, which is no longer co-linear with $z$ but now has the form $\textbf{k}_n = k_n (\cos \alpha \  \hat{\textbf{z}} - \sin \alpha \  \hat{\textbf{x}})$. Phase synchronicity is therefore accomplished if $\textbf{k}_n \cdot \hat{\textbf{z}} = \kappa c$. For a normal incidence of the laser ($\theta = \pi / 2$) this leads to the modified phase matching condition 
\begin{equation}
\lambda_p = \beta n \lambda \cos \alpha .
\label{eqn:rotatedphasematch}
\end{equation}
Hence as compared with the corresponding form from Eq. \ref{eqn:phasematch}, the grating period must be decreased by a factor $\cos \alpha$ in order to remain synchronous with the particle. This is geometrically obvious since in the rotated frame the apparent spacing between grating teeth is increased along $z$. Consequently, unless the periodicity is reduced, a particle of constant speed cannot remain in phase with the advancing phase fronts. If Eq. (\ref{eqn:rotatedphasematch}) is satisfied then for the case where the resonant mode is the fundamental ($n = 1$) the corresponding phase velocity now satisfies $\beta_n = \beta \cos \alpha$. This can be intuitively understood by noting that $\beta \cos \alpha$ is the component of the particle's velocity along the $z$-axis of the originally unrotated frame. 

We note that the $c_n^\pm$ coefficients may be different for the TE and TM modes and are complex-valued so may include a relative phase offset. For the case of normal incidence ($\theta = \pi/2$) and speed-of-light phase matching to the fundamental space harmonic ($n =1$, $\beta = 1$, $\lambda_p = \lambda \cos \alpha$) the vectors of the resonant mode take the forms
\begin{equation}
\textbf{Q}_n^{\text{TE}}=\left(
\begin{array}{c}
- \sin \alpha \tan \alpha \\
-i \sin \alpha \tan \alpha  \\
 \sin \alpha \\
\end{array}
\right), 
\textbf{Q}_n^{\text{TM}}=\left(
\begin{array}{c}
 - \sin \alpha \\
 -i \sin \alpha  \\
 \cos \alpha \\
\end{array}
\right) .
\label{eqn:Qsimple}
\end{equation}

\begin{figure*}
\centering
\includegraphics[width= 6in]{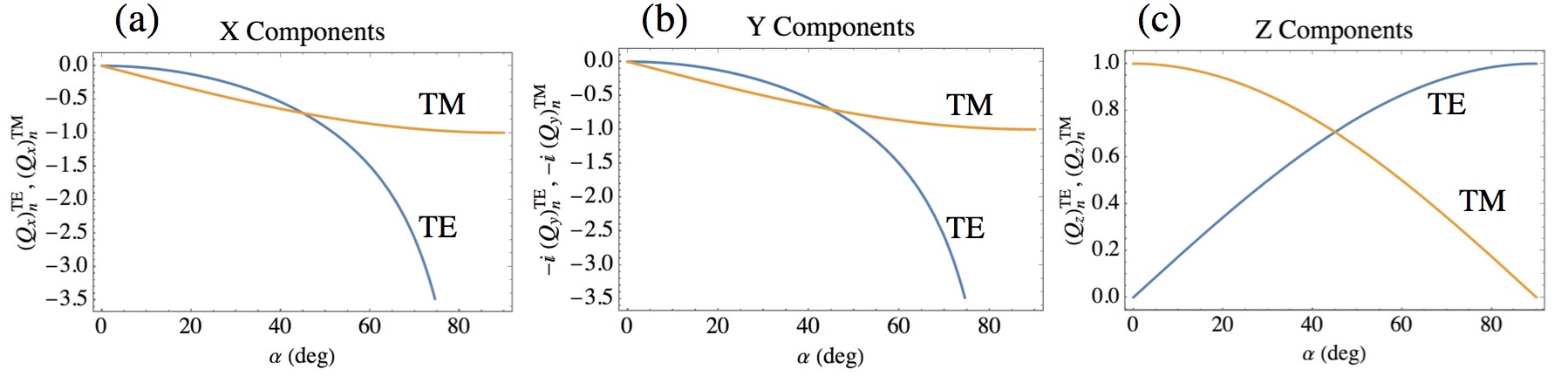}%
\caption{Plots of $\textbf{Q}_n$ coordinates in (a) $x$, (b) $y$, and (c) $z$ for the case $\beta = \beta_n = 1$, $\theta = \pi/2$ as functions of rotation angle $\alpha$.}%
\label{fig:Qplot}
\end{figure*}

The components of these vectors are plotted in Fig. \ref{fig:Qplot} as functions of $\alpha$. By rotating the polarization of the laser, as shown in Fig. \ref{fig:rotated}(a), we can form an arbitrary linear superposition of TE and TM modes ($\textbf{Q}_n^{\text{TE}} + \eta \textbf{Q}_n^{\text{TM}}$) where $\eta$ is a constant. From the forms of Eq. (\ref{eqn:Qsimple}) we see that $\textbf{Q}_n^\text{TE} = \textbf{Q}_n^\text{TM}  \tan \alpha $ and hence for the case of a 45 degree rotation ($\alpha = \pi / 4$), the two vectors are equal $\textbf{Q}_n^\text{TE} = \textbf{Q}_n^\text{TM}$ and give rise to the simple force relations
\begin{equation}
\begin{aligned}
&F_x=-q E_0 \frac{c_n^\pm}{\sqrt{2}} \left\{
\begin{array}{c}
 \cosh \left(\Gamma _ny\right) \\
 \sinh \left(\Gamma _ny\right) \\
\end{array}
\right\}e^{i \textbf{k}_n \cdot \textbf{r}}\text{  }  \\
&F_y=  \  q i E_0 \frac{c_n^\pm}{\sqrt{2}} \left\{
\begin{array}{c}
 \sinh \left(\Gamma _ny\right) \\
 \cosh \left(\Gamma _ny\right) \\
\end{array}
\right\}e^{i \textbf{k}_n \cdot \textbf{r}}\text{  } \\
&F_z=\ \  q E_0  \frac{c_n^\pm}{\sqrt{2}} \left\{
\begin{array}{c}
 \cosh \left(\Gamma _ny\right) \\
 \sinh \left(\Gamma _ny\right) \\
\end{array}
\right\}e^{i \textbf{k}_n \cdot \textbf{r}}\text{  }
\end{aligned}
\label{eqn:superforce}
\end{equation}
Hence, on the beam axis ($y = 0$), for incident lasers out-of-phase by $\pi$ [lower line of Eqs. (\ref{eqn:superforce})], the particle experiences a transverse deflection in the $y$ coordinate:
\begin{equation}
F_x = F_z = 0 \ , \ F_y = q \frac{i E_0}{\sqrt{2}} c_n^- e^{i \kappa z} \ 
\label{eqn:axialFy}
\end{equation}
For the case where the lasers are in-phase (upper line), the deflection is instead in the $x$ direction, but a nonzero axial longitudinal force arises: 
\begin{equation}
F_x = -q \frac{E_0}{\sqrt{2}} c_n^+ e^{i \kappa z} \ , \  F_y = 0 \ , \    F_z =  q \frac{E_0}{\sqrt{2}} c_n^+ e^{i \kappa z} 
\label{eqn:axialFxz}
\end{equation}

We see that the transverse fields no longer vanish in general for a synchronous speed of light particle except in the case $\alpha = 0$ consistent with the previous finding in the unrotated frame. The forces for the TE case appear to diverge when $\alpha \rightarrow \pi/2$. However this represents an inherently unphysical scenario, since the particle is then traveling parallel to the grating lines and so $\lambda_p \rightarrow 0$ in accordance with Eq. (\ref{eqn:rotatedphasematch}). 

\section{Discussion and Conclusion}
We've presented a consolidated formalism for the electric and magnetic forces in planar-symmetric structures with mirror symmetry about the $x$-$z$ plane and periodic in $z$ as shown in Fig. \ref{fig:dualstructure}. The symmetric axial deflecting modes represented in Eqs. (\ref{eqn:axialFy},\ref{eqn:axialFxz}) may be useful as steering elements for particle alignment in longer DLA structures, as an ultrafast temporal beam diagnostic (i.e. an optical transverse deflecting cavity), or as the basis for a laser-driven undulator \cite{plettner:2008,wootton:IPAC2015:deflection,england:FLS2018}. We note that for both the transverse deflection cases of Eqs. (\ref{eqn:axialFy},\ref{eqn:axialFxz}) and for longitudinal acceleration [TM case of Eq. (\ref{eqn:Qunrotated})], the sinh terms in Eq. (\ref{eqn:superforce}) result in a transverse force orthogonal to the deflection force that is either focusing or defocusing depending on the particle phase. Since this orthogonal force is out of phase with the deflection by $\pi/2$, it is zero for a particle sitting at the phase corresponding to maximum deflection. Bunches prepared by a previous DLA section or by an optical microbunching scheme such as that in Ref. \cite{sears:atto2008} will be bunched at the optical period of the laser and can therefore in principle be matched to the deflector in this way. 

For a resonant accelerating mode, longitudinally stable motion requires a choice of phase that is also transversely defocusing. For the case of a particle deflector, this constraint does not apply and so it should be feasible to phase the particles for peak deflection. However, since each microbunch still has a nonzero duration on the order of some fraction of a laser period, the front and the back of each microbunch will see strong focusing and defocusing forces. Hence, for a deflection element or undulator more than of order 1000 optical periods in length, some compensating force may be needed to improve transport and confine the beam in the corresponding coordinate. Compatible focusing techniques using the laser field itself have been proposed based on either alternating phase focusing or nonresonant harmonic focusing \cite{niedermayer:focusing:2018,naranjo_stable_2012}. A similar approach should be capable of compensating for defocusing forces associated with deflecting modes and should be the subject of further study. Since the deflection discussed here is electromagnetic, injection phase and bunching of the particle beam are critical. More detailed study of the particle dynamics, the resulting radiation field in a laser driven undulator, and plans for a demonstration experiment are underway and are discussed elsewhere in these proceedings \cite{ody:AAC2018}.

%I dunno about this one....
%Due to the nanoscopic size of the DLA period we are able to fit a significant amount of beam dynamics into a 2\,cm accelerator. In terms of the time-averaged particle motion the beam undergoes about 4.6 synchrotron oscillations and 4.4 betatron oscillations. In this, or in slightly longer DLA's, we have the possibility of study tune resonances in a strongly non-linear, coupled accelerator. Additionally,  we have only considered a perfectly symmetric TM accelerating mode, but it is easy to make a DLA with a time-varying dipole moment which could be used to drive instabilities in the focusing channel. This makes DLA a compact platform for.... 

\section*{Acknowledgments}
This work was supported by Gordon and Betty Moore Foundation (GBMF4744), National Science Foundation (NSF) (PHY-1734215, PHY-1535711), and U.S. Department of Energy (DE-AC02-76SF00515, DE-SC0009914).

\bibliographystyle{IEEEtran}
\bibliography{england_aac18}


%\bibliographystyle{IEEEtran}

%\begin{thebibliography}{1}


%\end{thebibliography}

\end{document}